# Dynamics of a capacitive electret-based microcantilever for energy harvesting

Mahyar Ghavami[1], Saber Azizi[2*], Mohammad Reza Ghazavi[1]

[1]Tarbiat Modares University, Tehran, Iran

[2]Urmia Univesity of Technology, Urmia, Iran

## Abstract

In this paper, a novel electret-based capacitive energy harvesting device has been developed according to out-of-plane gap closing scheme. The device is composed of a micro cantilever and a substrate which form a variable capacitor and is in series with a resistance. An electret material is used to provide the bias voltage which is needed in capacitive energy harvesters in order to scavenge energy from ambient vibration. The ambient vibration is applied to the system as a harmonic base excitation. The motion equations and the corresponding boundary conditions are derived using Hamilton' principle based on Euler-Bernoulli beam theory and the Kirchhoff's voltage law is employed to couple the mechanical and electrical fields. The equations of motion are discretized using Galerkin procedure and integrated numerically over time. Pull-in instability of the system is investigated in both static and dynamic cases. The effect of various parameters on the behavior of the device is studied. The maximum theoretical harvested power is resistance in the order of $1\ \mu W$.

## 1. Introduction

Energy harvesting is to scavenge or harvest energy from a variety of ambient energy sources (e.g. solar, wind and vibration) and convert it into electrical energy for self-powered wireless electronic applications ranging from structural health monitoring to medical implants [1-3]. Vibration energy harvesting devices are developed based on three basic energy conversion mechanisms including, electromagnetic [4, 5], piezoelectric [6-8] and electrostatic (capacitive) [9-12] transductions. Since the technology for manufacturing electrostatic transducers such as capacitive-based sensors and actuators (e.g., accelerometers, gyroscopes, comb drives) is well established, it is beneficial to utilize the same technology (standard MEMS technology) for manufacturing capacitive energy harvesting devices. This allows the production of devices in large numbers at low cost [13]. The basic of electrostatic energy harvesting devices is a charged variable capacitor which its capacitance changes as a result of ambient vibrations. Any change in the capacitance leads to a charge rearrangement on the capacitor electrodes and consequently the charge flows through the electrical circuit. Based on how the capacitance changes, capacitive energy harvesting devices are classified into three different types, including in-plane overlap which varies the overlap area between electrodes, in-plane gap closing which varies the gap between electrode fingers, and out-of-plane gap closing which varies the gap between two electrodes [2]. These devices can operate in either switched or continuous scheme. In switched systems, some controlled switches are employed to change the capacitor's condition discontinuously through voltage-constrained or charge-constrained cycles, however, continuous systems do not require any controlled switches to operate and any change in the biased capacitor causes a charge flow on the resistance. Since the utilization of switches requires some extra circuitry to control them and precious energy is consumed by that control circuitry, hence, continuous systems are more advantageous than switched ones [2]. Electrostatic generators require an initial voltage to charge the capacitor and initiate the energy conversion. This bias voltage can be provided by electret materials in continuous systems. Electrets are dielectric materials with quasi-permanent charge which can provide an electric field for tens of years [14]. Various materials have been used as electrets in electrostatic energy harvesters such as Teflon [15], $SiO_2$ [11] and CYTOP [16]. Among these materials, CYTOP is reported to have a high charge density and is a MEMS-compatible perfluoropolymer [16].

Many researchers have studied the electret-based energy harvesters. Boisseau et al [9] developed an out-of-plane energy harvester in which a cantilever beam with a tip mass is used as the movable electrode of the variable capacitor and an electret layer is employed to provide the bias voltage. The resonant system is modeled as a spring-mass to simulate the mechanical behavior of the device. Tsutsumino et al [16] developed an in-plane overlap energy harvester where CYTOP is used as a high performance electret material to supply the bias voltage. The resonant system consist of a proof mass with parylene high-aspect-ratio springs which allows large amplitude oscillation and low response frequency. Tao et al [14] proposed an out-of-plane energy harvesting device with dual-charged electret where the resonant system consists of a movable circular mass with a series of spiral spring and two LDPE electret plates with positive and negative charge are used as the voltage source. Wang et al [17] presented an out-of-plane gap electrostatic energy harvester with a proof mass suspended by beam type springs. The device consists of a four wafer stack where CYTOP is employed both as the electret material and adhesive interface between wafers. Peano et al [18] introduced and optimized a capacitive electret-based energy harvesting device with in-plane overlap scheme using a nonlinear dynamical model. Sterken et al [19] proposed, optimized and fabricated an electret micro generator with in-plane overlap scheme. The device consist of two bonded wafers as the electrodes of the variable capacitor. The system is analyzed using a lumped element model based on the equivalency between mechanical and electrical elements. Tvedt et al [20] introduced an in-plane overlap energy harvesting device with an electret material to polarize the electrodes. The resonant system is modeled as a lumped equivalent circuit based on both linear and nonlinear models. Halvorsen et al [21] designed and fabricated an in-plane overlap electret-based energy harvesting device. The resonant system consist of a proof mass supported by springs above an oxidized silicon substrate with patterned electret stripes. Kloub et al [22] optimized an in-plane overlap micro capacitive energy harvesting device by maximizing the capacitance variation. The device is composed of a seismic mass suspended by beams and modeled as a spring-mass-damper system. Miki et al [23] developed and fabricated an in-plane electret energy harvester with a proof mass suspended by hybrid high-aspect ratio parylene springs. Genter et al [24] fabricated an out-of-plane electret energy harvester in which Parylene-C is used as eletret layer and spring material. The device is composed of two silicon wafers as base chip and resonator chip with a proof mass suspended by a flexural spring from the resonator chip.

To the author's knowledge, most researches on the electret-based energy harvesters have modeled the resonant system as a mass-spring model and mostly focused on the electrical behavior of the device.

In this paper an electret-based energy harvester with out-of-plane gap closing scheme is proposed where a micro cantilever is used as the moving electrode of the variable capacitor. There are a few researches who proposed a cantilever-based energy harvester and they have modeled the vibrating system as a spring-mass [9]. In this study, the microbeam is modeled as a continuous system based on Euler-Bernoulli beam theory and the characteristics of the device are investigated with a focus on the mechanical behavior of the energy harvester. An electret layer is attached to stationary electrode (substrate) to supply the bias voltage. This variable capacitor is in series with a resistance which consumes the harvested energy. The simplicity of the presented device enables a low-cost manufacturing process with MEMS technology.

## 2. Modeling

A schematic representation of the proposed energy harvester is shown in Fig. 1. The device is composed of a variable capacitance which is formed by a substrate and a micro cantilever of length $L$, width $b$ and thickness $h$. Since capacitive energy harvesters require a voltage source to scavenge energy from the ambient vibration, therefore, an electret layer is attached to substrate to provide the bias voltage. This capacitor is in series with a resistance in an electric circuit to consume the harvested energy. The complex is mounted on a package and the ambient vibration is applied in terms of a harmonic base excitation. The coordinate system is originated on the microbeam at its clamped end and is considered such that the *xoy* plane is coincident with midplane of the undeformed configuration. The x axis coincide with the centroidal axis of the microbeam and the positive direction of z axis is downwards. In order to model the mechanical behavior of the micro cantilever, the Euler-Bernoulli beam theory is employed which neglects the shear deformation and rotary inertia effects. In order to harvest the maximum energy, the system is assumed to work in vacuum medium and the effects of damping forces are neglected.

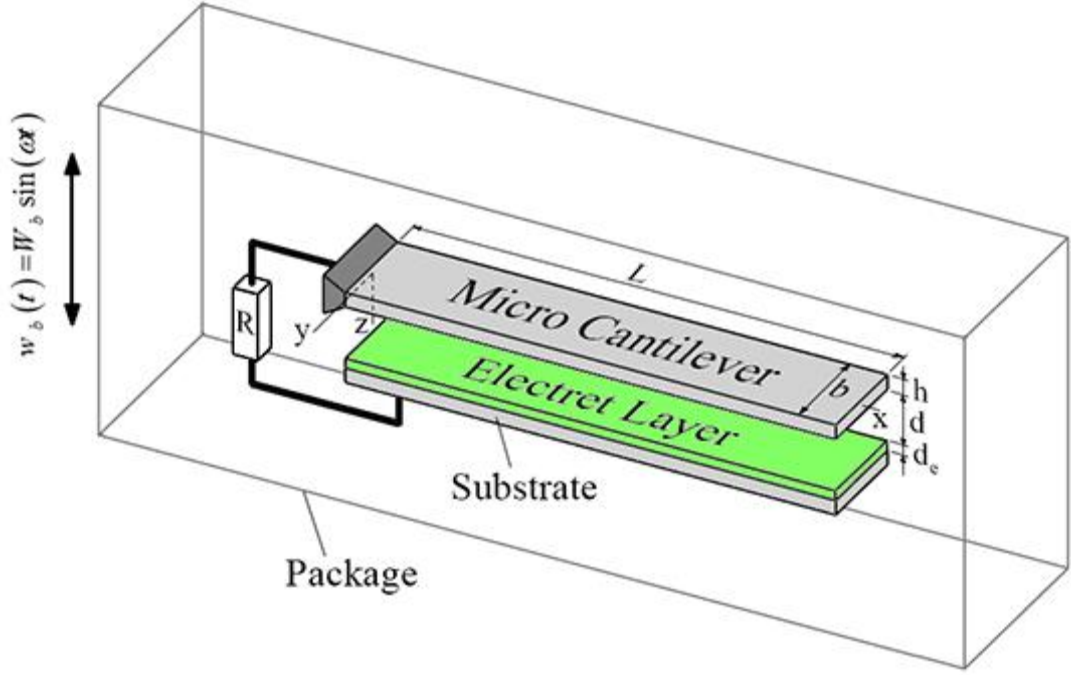

**Figure 1.** Schematic representation of the energy harvesting device

The motion equation of the microbeam is derived using Hamilton's principle which is given by [25]:

$$\int_{t_1}^{t_2} (\delta T - \delta U )dt = 0 \qquad (1)$$

where $T$ and $U$ are kinetic and potential energies of the system, respectively. The potential energy of the system is composed of mechanical and electrical parts as:

$$U = U_m + U_E \qquad (2)$$

where $U_m$ and $U_E$ are mechanical and electrical potential energies, respectively.

The strain energy of the microbeam based on Euler-Bernoulli beam theory will be obtained as follow [25]:

$$U_m = \frac{1}{2}\int_0^L EI \left[ \frac{\partial^2 w(x,t)}{\partial x^2} \right]^2 dx \qquad (3)$$

where $E$ is Young's modulus, $A$ is cross sectional area and $I$ is the moment of inertia of the microbeam with respect to its neutral axis.

The electrical potential energy is equal to potential energy of the capacitor and is expressed as [9]:

$$U_E = \frac{Q^2(t)}{2C(t)} \tag{4}$$

where $Q$ is induced electric charge on the upper electrode (micro cantilever) and $C$ denotes the equivalent capacitance of the system which consists of two series capacitances including the electret layer capacitance and the variable capacitance. To obtain the equivalent capacitance of the device, the capacitor is considered as an infinite number of differential parallel capacitors, each of them are in series with the electret layer capacitance. Neglecting the fringing effects and assuming complete overlap between the micro cantilever and the substrate, the equivalent capacitance of the system is obtained as follow:

$$C(t) = \int_0^L \frac{\varepsilon_0 b \, dx}{d - w(x,t) + \frac{d_e}{\varepsilon_e}} \tag{5}$$

where $d$ is the gap between the upper electrode and the electret layer, $d_e$ is the electret layer thickness, $\varepsilon_e$ is relative dielectric constant of electret layer and $\varepsilon_0$ is dielectric constant of vacuum.

The kinetic energy of the microbeam is given as:

$$T = \frac{1}{2} \int_0^L \rho A \left[ \frac{\partial w(x,t)}{\partial t} + \frac{dw_b(t)}{dt} \right]^2 dx \tag{6}$$

where $\rho$ is the mass density of the microbeam and $w_b$ is the base displacement due to the external excitation which is considered as a harmonic motion as follows:

$$w_b(t) = W_b \sin(\omega t) \tag{7}$$

in which $W_b$ and $\omega$ are the amplitude and frequency of the ambient vibrations, respectively.

Substituting Eqs. (2-7) into Hamiltonian, integrating by parts with respect to $t$ and $x$ and applying fundamental lemma of calculus, the equation of motion and the associated boundary conditions of the micro cantilever are obtained as:

$$EI \frac{\partial^4 w(x,t)}{\partial x^4} + \rho A \frac{\partial^2 w(x,t)}{\partial t^2} = \frac{\varepsilon_0 b Q^2(t)}{2C^2(t)\left[d - w(x,t) + d_e/\varepsilon_e\right]^2} - \rho A \frac{d^2 w_b(t)}{dt^2} \tag{8}$$

$$\begin{cases} x = 0: & w = 0, \quad \frac{\partial w}{\partial x} = 0 \\ x = L: & \frac{\partial^2 w}{\partial x^2} = 0, \quad \frac{\partial^3 w}{\partial x^3} = 0 \end{cases} \tag{9}$$

In order to couple the governing equations of mechanical and electrical fields, the Kirchhoff's voltage law is employed. Neglecting the parasitic capacitances, the electric circuit of the energy harvester is modeled as depicted in Fig. 2.

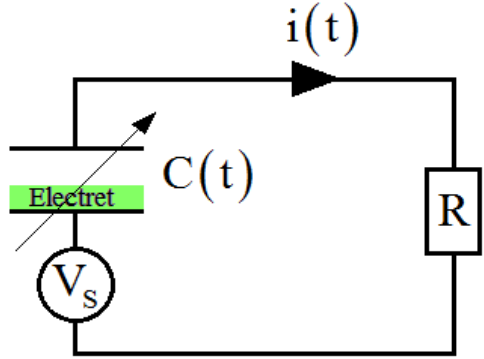

**Figure 2.** The equivalent electrical circuit of the device

Therefore, one can obtain the governing electrical equation of the system by applying Kirchhoff's voltage to the equivalent circuit, as follow:

$$\frac{dQ(t)}{dt} = \frac{V_S}{R} - \frac{Q(t)}{RC(t)} \tag{10}$$

where $R$ and $V_S$ are resistance and surface voltage of the electret layer, respectively.

To obtain the dimensionless form of the governing equations, the following dimensionless parameters are introduced:

$$\hat{x} = \frac{x}{L} \quad, \quad \hat{w} = \frac{w}{d} \quad, \quad \hat{w}_b = \frac{w}{d} \quad, \quad \hat{Q} = \frac{Q}{Q_e}$$

$$\hat{W}_b = \frac{W_b}{d}\Omega^2 \quad, \quad \hat{t} = t\sqrt{\frac{EI}{\rho A L^4}} \quad, \quad \Omega = \omega\sqrt{\frac{\rho A L^4}{EI}} \tag{11}$$

where $Q_e$ is the electric charge stored in electret layer and equals to $Q_e = C_e V_S$, in which $C_e$ is the electret layer capacitance and is given by [26]:

$$C_e = \frac{\varepsilon_0 \varepsilon_e b L}{d_e} \tag{12}$$

Substituting the dimensionless parameters into Eq. (8) and Eq. (10) and removing the hat notation for brevity, the nondimensional governing equations are obtained as follow:

(13-a)

$$\begin{cases} \dfrac{\partial^4 w(x,t)}{\partial x^4} + \dfrac{\partial^2 w(x,t)}{\partial t^2} = \dfrac{\alpha_1 V_S^2}{k-1} \dfrac{Q^2(t)}{\left[k-w(x,t)\right]^2 \left[\int_0^1 \dfrac{dx}{k-w(x,t)}\right]^2} + W_b \sin(\Omega t) \\ \dfrac{dQ}{dt} + \alpha_2 \dfrac{Q}{\int_0^1 \dfrac{dx}{k-w(x,t)}} = \alpha_3 \end{cases} \quad (13\text{-a})$$

where

$$\alpha_1 = \dfrac{\varepsilon_0 b L^4}{2 E I d^3} \quad , \quad \alpha_2 = \dfrac{d}{R \varepsilon_0 b L} \sqrt{\dfrac{\rho A L^4}{EI}} \quad , \quad \alpha_3 = \dfrac{1}{R C_e} \sqrt{\dfrac{\rho A L^4}{EI}} \quad , \quad k = 1 + \dfrac{d_e}{\varepsilon_e d} \quad (14)$$

The dimensionless boundary conditions reduce to:

$$\begin{cases} w = 0 \quad , \quad \dfrac{\partial w}{\partial x} = 0 \quad \text{at} \quad x = 0 \\ \dfrac{\partial^2 w}{\partial x^2} = 0 \quad , \quad \dfrac{\partial^3 w}{\partial x^3} = 0 \quad \text{at} \quad x = 1 \end{cases} \quad (15)$$

## 3. Solution procedure

The Galerkin method is employed to discretize the integro-partial differential equation of the system. To this end, the transverse deflection of the microbeam is considered in the form of the following expansion series:

$$w(x,t) = \sum_{j=1}^{N} q_j(t) \varphi_j(x) \quad (16)$$

where $N$ is number of modes, $q_j(t)$ is the generalized coordinate of $j$th mode and $\varphi_j(x)$ is $j$th mode shape of the microcantilever in free vibration. Multiplying Eq. (13-a) by $\left[k - w(x,t)\right]^2$, inserting Eq. (16) into resulting expression, multiplying it in $\varphi_i(x)$ and integrating over dimensionless length of the microbeam, the reduced order model of the system reduces to:

$$k^2 \sum_{j=1}^{N} q_j \int_0^1 \varphi_i \varphi_j^{IV} dx - 2k \sum_{j=1}^{N} \sum_{k=1}^{N} q_j q_k \int_0^1 \varphi_i \varphi_j \varphi_k^{IV} dx$$

$$+ \sum_{j=1}^{N} \sum_{k=1}^{N} \sum_{l=1}^{N} q_j q_k q_l \int_0^1 \varphi_i \varphi_j \varphi_k \varphi_l^{IV} dx + k^2 \sum_{j=1}^{N} \ddot{q}_j \int_0^1 \varphi_i \varphi_j dx$$

$$- 2k \sum_{j=1}^{N} \sum_{k=1}^{N} \ddot{q}_j q_k \int_0^1 \varphi_i \varphi_j \varphi_k dx + \sum_{j=1}^{N} \sum_{k=1}^{N} \sum_{l=1}^{N} q_j q_k \ddot{q}_l \int_0^1 \varphi_i \varphi_j \varphi_k \varphi_l dx$$

$$+ k^2 c_d \sum_{j=1}^{N} \dot{q}_j \int_0^1 \varphi_i \varphi_j dx - 2k c_d \sum_{j=1}^{N} \sum_{k=1}^{N} \dot{q}_j q_k \int_0^1 \varphi_i \varphi_j \varphi_k dx$$

$$+ c_d \sum_{j=1}^{N} \sum_{k=1}^{N} \sum_{l=1}^{N} q_j q_k \dot{q}_l \int_0^1 \varphi_i \varphi_j \varphi_k \varphi_l dx \qquad (17)$$

$$= \frac{\alpha_1 V_S^2}{k-1} \frac{Q^2}{\left( \int_0^1 \frac{dx}{K - \sum_{j=1}^{N} q_j \varphi_j} \right)^2} \int_0^1 \varphi_i dx + k^2 W_b \sin(\Omega t) \int_0^1 \varphi_i dx$$

$$- 2k W_b \sin(\Omega t) \sum_{j=1}^{N} q_j \int_0^1 \varphi_i \varphi_j dx$$

$$+ W_b \sin(\Omega t) \sum_{j=1}^{N} \sum_{k=1}^{N} q_j q_k \int_0^1 \varphi_i \varphi_j \varphi_k dx \qquad , \qquad N = 1, 2, 3, ...$$

$$\frac{dQ}{dt} + \alpha_2 \frac{Q}{\int_0^1 \frac{dx}{k - \sum_{j=1}^{N} q_j \varphi_j}} = \alpha_3 \qquad , \qquad N = 1, 2, 3, ... \qquad (18)$$

In order to obtain the dynamic characteristics of the energy harvester, a single degree of freedom model is considered and the differential equations are integrated numerically to attain the dynamic response of the device. The performance of the energy harvester is evaluated by calculating the mean output power based on steady state response as [9]:

$$\bar{P} = \frac{1}{t_2 - t_1} \int_{t_1}^{t_2} R \left( \frac{dQ}{dt} \right)^2 dt \qquad (19)$$

To calculate the static pull-in voltage of the device and static position of the microbeam, the time-dependent terms in Eq. (13-a) are dropped and static configuration of the microbeam is determined based on:

$$\frac{d^4 w}{dx^4} = \alpha_1 \frac{V_S^2}{(k-w)^2} \qquad (20)$$

By applying the Galerkin procedure in Eq. (20), the pull-in voltage and the equilibrium position of the system can be determined for static case.

## 4. Results and discussion

The mechanical, electrical and geometrical properties of the studied model are given in Table 1. CYTOP® is used as the electret material which has a break down voltage of 90 kV/mm [27].

**Table 1.** System Properties

| Properties | Notation | Magnitude |
|---|---|---|
| Microbeam length | $L$ | 100 μm |
| Microbeam width | $b$ | 20 μm |
| Microbeam thickness | $h$ | 4 μm |
| Young's modulus[28] | $E$ | 169.2 GPa |
| Shear modulus [28] | $\mu$ | 65.8 GPa |
| Gap | $d$ | 3 μm |
| Electret layer thickness | $d_e$ | 2 μm |
| Surface voltage of electret layer | $V_s$ | 180 V |
| Relative permittivity of electret layer [27] | $\varepsilon_e$ | 2 |

The static pull-in voltage of the device is determined using Eq. (20) and compared with that of dynamic case (Fig. 3). The dynamic pull-in voltage of the system is calculated by numerically solving the governing motion equation (Eq. (13-a)) in the absence of the resistance and external excitation. As expected, the dynamic pull-in voltage is lower than the static one. The dynamic pull-in voltage determines the charging limit of the electret layer which should not be exceed to prevent the occurrence of pull-in phenomenon in the device.

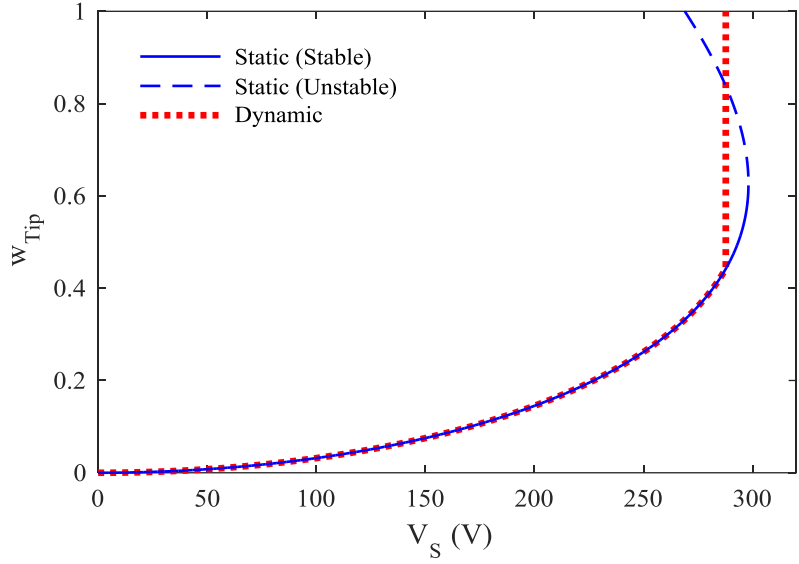

**Figure 3.** Maximum dimensionless tip deflection versus electret surface voltage in static and dynamic case

In order to assure the accuracy of the electro-mechanical model, an energy conservation study is performed on the device. To this end, an initial displacement is given to the microbeam's tip in the absence of base excitation. The initial energies, including the strain energy stored in the deflected microbeam and electrical energy stored in the capacitor, and the harvested energy which is consumed in the resistance, are numerically determined. To satisfy the energy conservation law, the harvested energy needs to be equal to the initial mechanical work performed on the microbeam. The results of energy conservation study are represented in Table 2 for various resistances and initial deflections of the microcantilever's tip deflection. It can be seen that the difference between initial and harvested energies is lower than 3 percent for tip's displacement lower than 0.2 times the gap and as it could be expected, the harvested energy is lower than initial mechanical work injected to the system.

Also this error grows as the initial deflection of the microbeam' tip increases. This is attributed to the fact that the linear classical beam theory is valid for small ratios of $w/h$ and as this ratio increases, the shortening effect which is neglected in classical beam theory, must be taken into account to obtain more precise results.

**Table 2.** The difference between the injected and the harvested energies in the absence of external excitation ($V_s = 180$ V)

| $R\ (\Omega)$ | $w_{Initial}(1)/d$ | | | | | | | |
| --- | --- | --- | --- | --- | --- | --- | --- | --- |
| | **0.13** | **0.14** | **0.15** | **0.16** | **0.17** | **0.18** | **0.19** | **0.2** |
| $10^7$ | -0.466 % | -0.774 % | -1.08 % | -1.385 % | -1.688 % | -1.991 % | -2.291% | -2.59 % |
| $10^8$ | -0.466 % | -0.774% | -1.08 % | -1.385 % | -1.688 % | -1.991 % | -2.291 % | -2.59 % |
| $2\times 10^8$ | -0.466 % | -0.774 % | -1.08 % | -1.385 % | -1.688 % | -1.991 % | -2.291 % | -2.59 % |
| $3\times 10^8$ | -0.466 % | -0.774 % | -1.08 % | -1.385 % | -1.688 % | -1.991 % | -2.291 % | -2.59 % |
| $4 \times 10^8$ | -0.466 % | -0.774 % | -1.08 % | -1.385 % | -1.689 % | -1.991 % | -2.291 % | -2.59 % |
| $5 \times 10^8$ | -0.466 % | -0.774 % | -1.08 % | -1.385 % | -1.689 % | -1.991 % | -2.291 % | -2.59 % |

The temporal response of the system in free oscillations is illustrated in Fig. 4 for a resistance of $R = 100 \text{ M}\Omega$ and the initial tip displacement $w_{Initial}(1)/d = 0.15$.

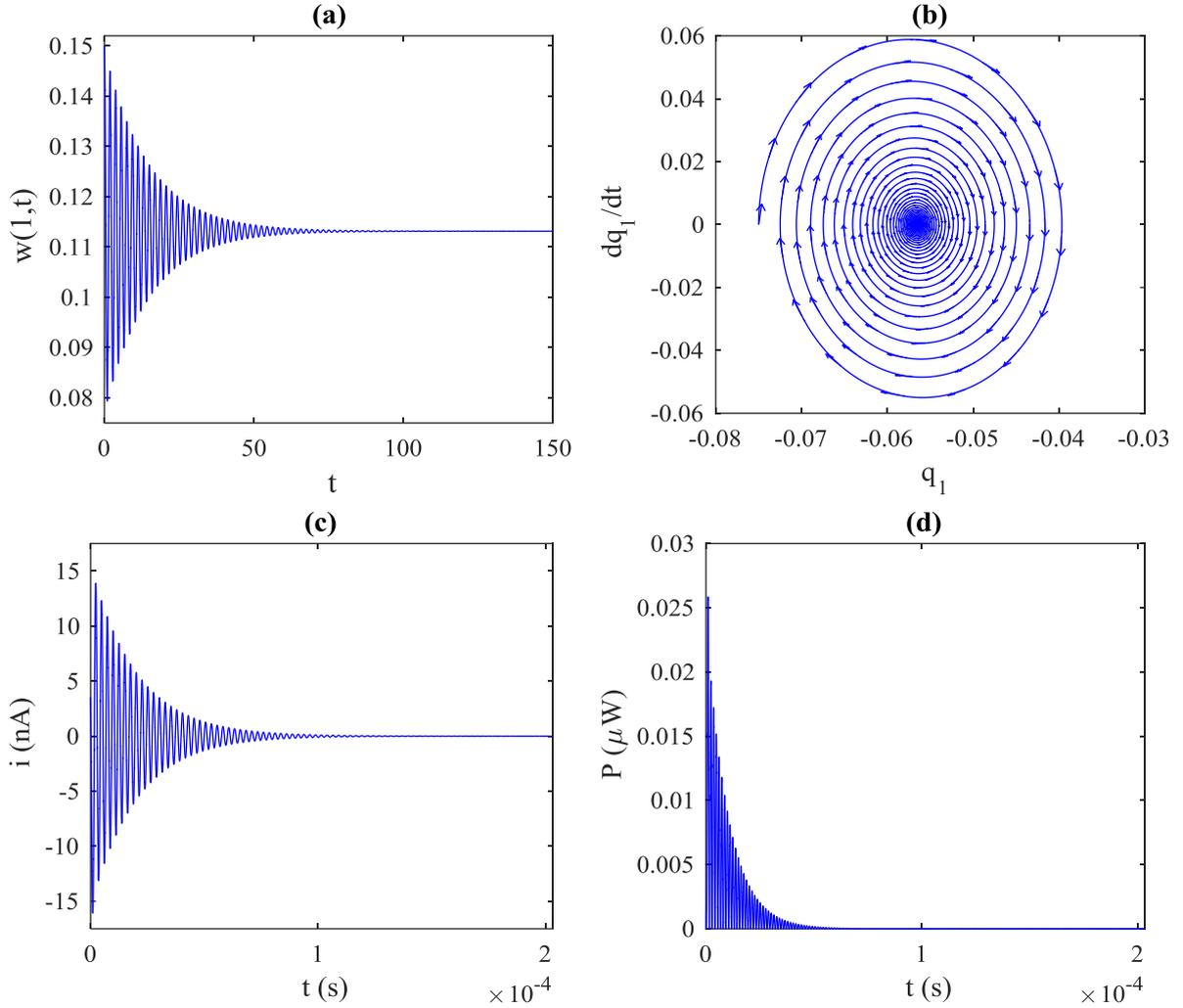

**Figure 4.** Free vibration response of the system for $R = 100 \text{ }\Omega$, $V_S = 180 \text{ V}$ and initial tip displacement $w_{Initial}(1)/d = 0.15$; (a) Dimensionless deflection of microbeam's tip, (b) Phase plane, (c) Electric current in the resistance, (d) Output power

As depicted in Fig. 4, the resistance consumes the energy which is injected to the system and gradually and accordingly due to the reduction of the total mechanical energy, the deflection of the microbeam gradually decreases. This effect of the **resistance** resembles a damping force that acts on the microbeam. To determine the equivalent damping ratio of the corresponding resistances, a curve fitting approach is employed to obtain the envelope equation of temporal response of the microbeam in free oscillations. The variation of nondimensional equivalent damping ratios with resistance is presented in Fig. 5 for different surface voltage of the electret layer. It is concluded that the equivalent damping ratio increases as the resistance increases. This behavior continues up to a critical resistance- approximately $R = 100$ M$\Omega$ more than which the equivalent damping coefficient decreases this is known as Lorenzian behavior in the literature [6]. The equivalent damping ratio resembles the rate in which the energy is harvested by device, therefore, in the free vibration case, the system has maximum output power for $R = 100$ M$\Omega$ resistance. Also we concluded that the equivalent damping ratio and consequently the output power of the device increases by increasing the surface voltage of the electret layer.

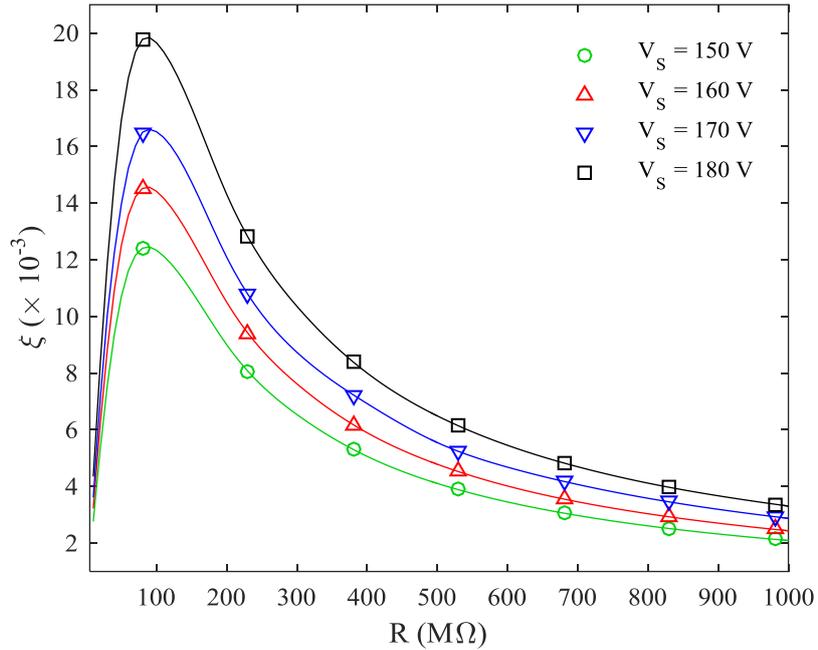

**Figure 5.** Variation of nondimensional equivalent damping ratio with resistance for different value of electret surface voltage

In order to evaluate the performance of the energy harvester in the presence of base excitation which is induced by ambient vibrations, the frequency response curve of the micro cantilever and also the mean output power of the device are obtained for different values of resistance. The frequency response curve of the microbeam shown in Fig. 6, for various resistances with an electret surface voltage of $V_S = 180$ V. Fig. 7 illustrates the output power of the energy harvester as a function of excitation frequency for various resistances.

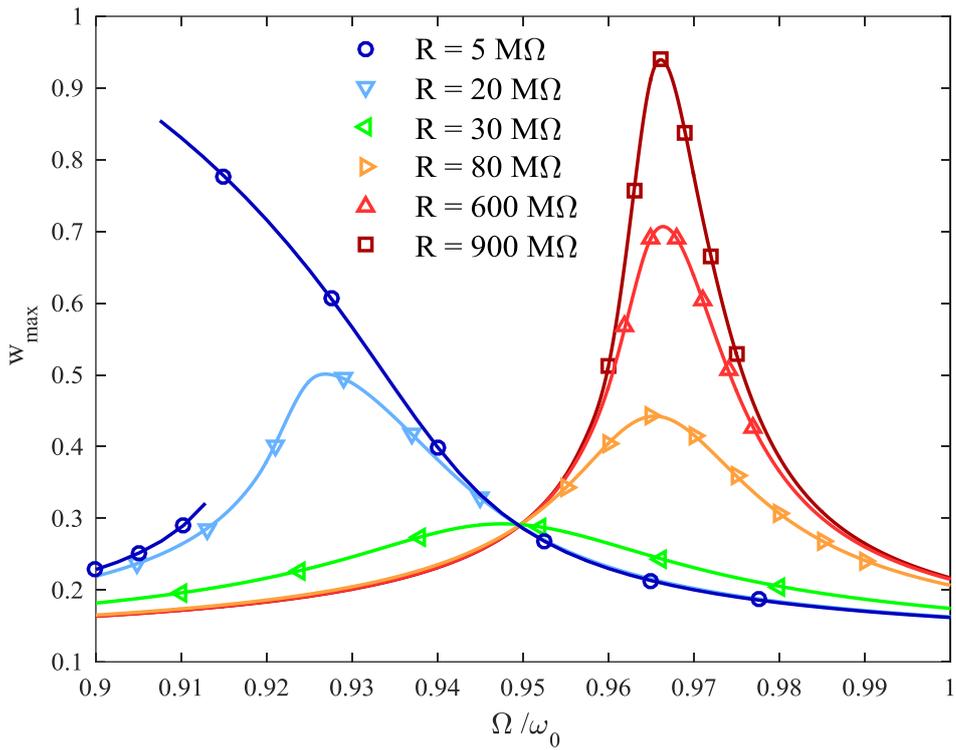

**Figure 6.** Frequency response curve for various resistances for $V_S = 180$ V and $W_b = 0.05$

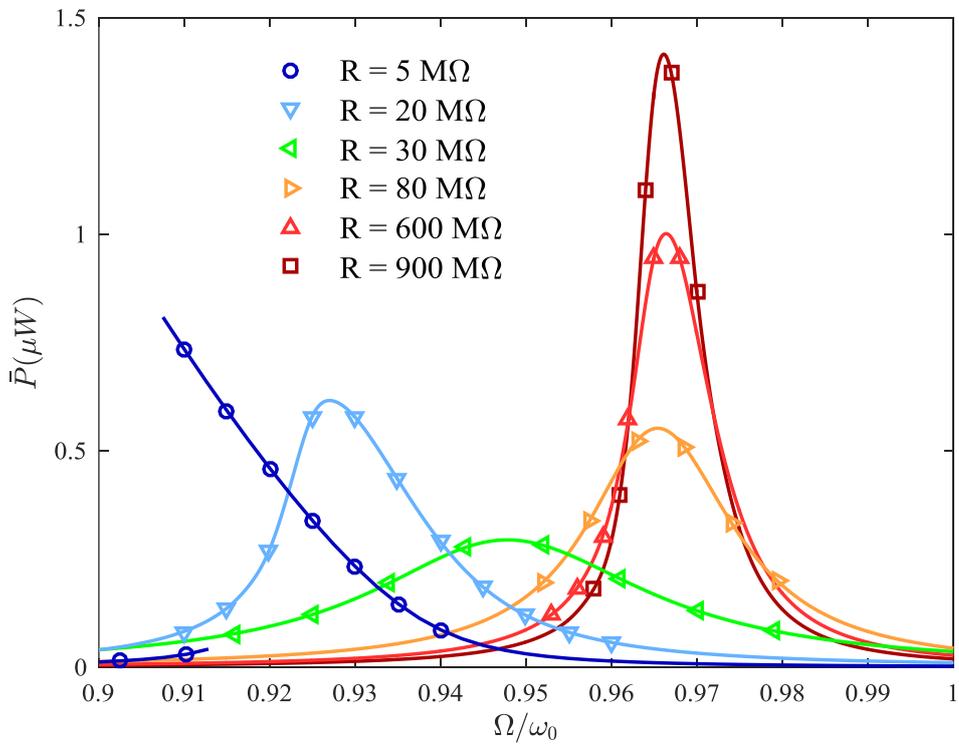

**Figure 7.** Frequency response curve for various resistances for $V_S = 180$ V and $W_b = 0.05$

In low resistances, a relatively large electrostatic pressure acts between the microbeam and the substrate and hence it reduces the microbeam's stiffness [29]. Therefore, in this case the microbeam exhibits a softening behavior and the maximum amplitude and consequently, the maximum output power occurs in low frequencies Due to this softening behavior, the microbeam has multiple responses in some range of excitation frequency and the response depends on the initial condition of the system in this region. By increasing the resistance, the voltage between the variable capacitor electrodes decreases in resonance region and therefore, due to a lower electrostatic force between the microbeam and the stationary electrode, the stiffness of the microbeam increases (the softening effect weakens and eventually vanishes) which leads the resonance frequency of the system to shift toward higher frequencies. Consequently, the jump phenomenon disappears and the microbeam undergoes a linear type response. As illustrated in Fig. 6, by increasing the resistance, the resonance frequency shifting decreases gradually and finally the resonance frequency remains almost constant and the dynamic deflection starts to grow. In this case, further the resistance increases, further the voltage of the capacitor decreases and consequently, the dynamic deflection of the microbeam increases until in large resistances the microbeam contacts with the electret layer and the pull-in occurs. It should be noted that as mentioned earlier, the resistance has a damping effect on the micro cantilever which causes the resonance to occur in frequencies lower than the natural frequency of the system. The variation of the microbeam's tip deflection with resistance and excitation frequency is shown in Fig. 8 for a surface voltage of $V_S = 180$ V and non-dimensional excitation amplitude of $W_b = 0.05$.

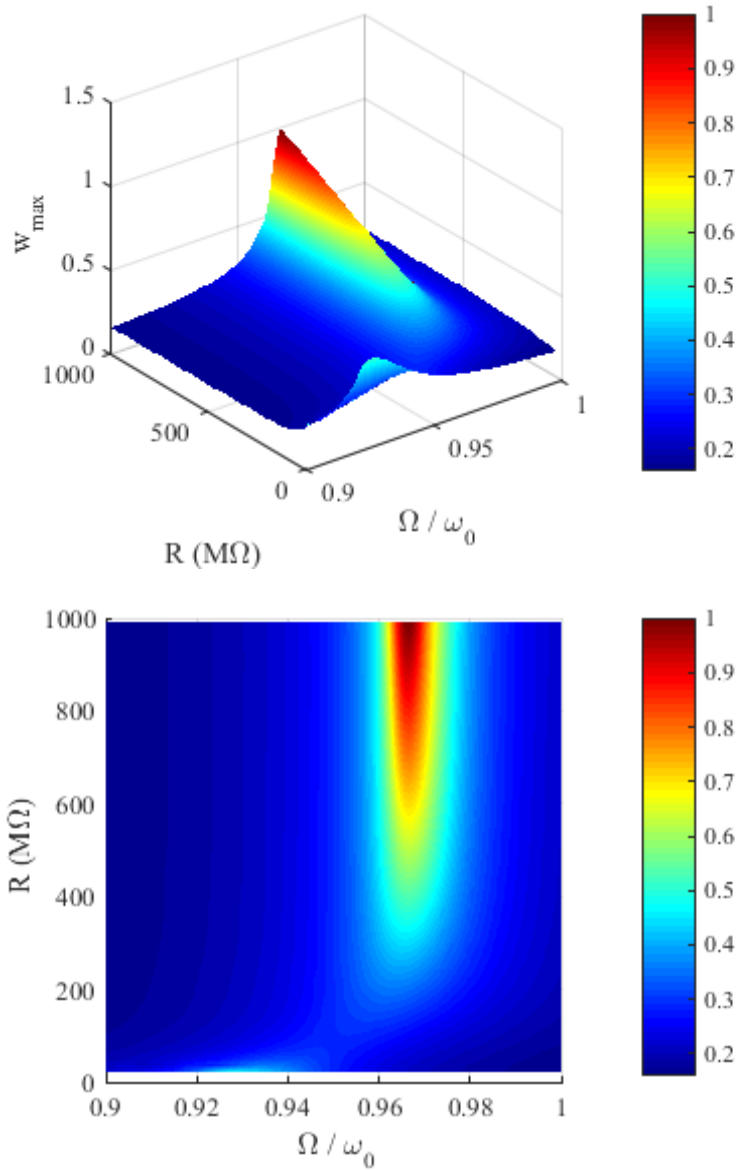

**Figure 8.** Maximum nondimensional deflection of the microbeam as a function of excitation frequency and resistance for $V_S = 180$ V and $W_b = 0.05$

Fig. 9 shows variation of the mean output power of the micro harvester with resistance and excitation frequency of $V_S = 180$ V and $W_b = 0.05$, respectively. The mean output power has a direct relationship with microbeam's deflection. Therefore, the mean output power of the device resembles the response of the microbeam. In very low and very high excitation frequencies, the mean output power of the device is similar to the response of the system in free oscillations. It can be seen that in some range of resistance and ambient vibration frequencies, the device delivers a mean output power on the order of 1 microwatt.

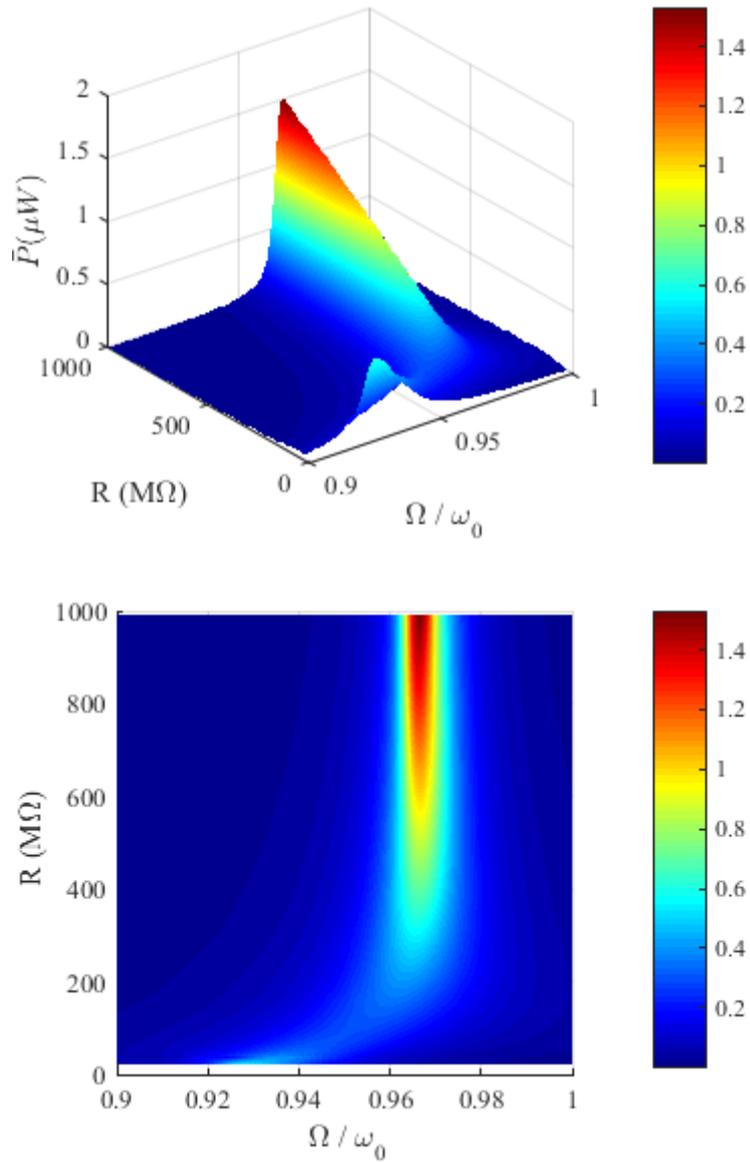

**Figure 9.** Mean output power of the energy harvester as a function of excitation frequency and resistance for $V_S = 180$ V and $W_b = 0.05$

Fig. 10 shows the variation of mean output power of the system with resistance in excitation frequencies near the resonance frequency of lower and higher resistances.

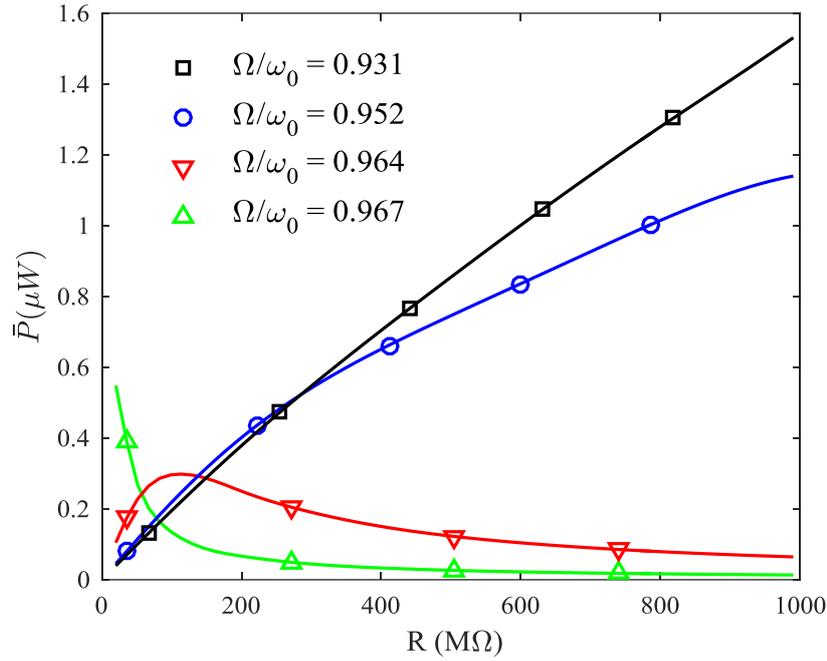

**Figure 10.** The variation of mean output power with resistance as a function of excitation frequency for $V_S = 180$ V and $W_b = 0.05$

In excitation frequencies near the resonance frequency of lower resistances $(0.925 < \Omega/\omega_0 < 0.935)$, the low resistance leads to a large electric current in the circuit and it causes the mean output power to be relatively high. As the resistance increases in these frequencies, the electric current decrease and yields the mean output power to decrease, however the resistance has increased; therefore it can be concluded that the electric current has a dominating effect on variation of the mean output power with resistance in the aforementioned frequency domain. By increasing the excitation frequency, the dominancy of the electric current on the mean output power subsides and the resistance becomes more prevailing. In excitation frequencies near the resonance frequencies of higher resistances $(0.96 < \Omega/\omega_0 < 0.97)$, the variation of mean output power of the system with resistance, is mostly affected by the resistance. That is to say, by increasing the resistance in these excitation frequencies, however the electric current is low, but the mean output power of the energy harvester increases due to the resistance dominancy. This behavior is holds for frequency response of the microbeam. The variation of maximum electric current with the resistance and the excitation frequency is shown in Fig. 11 for $V_S = 180$ V; in low resistances, the electric current in resonance region is larger than that of high resistances.

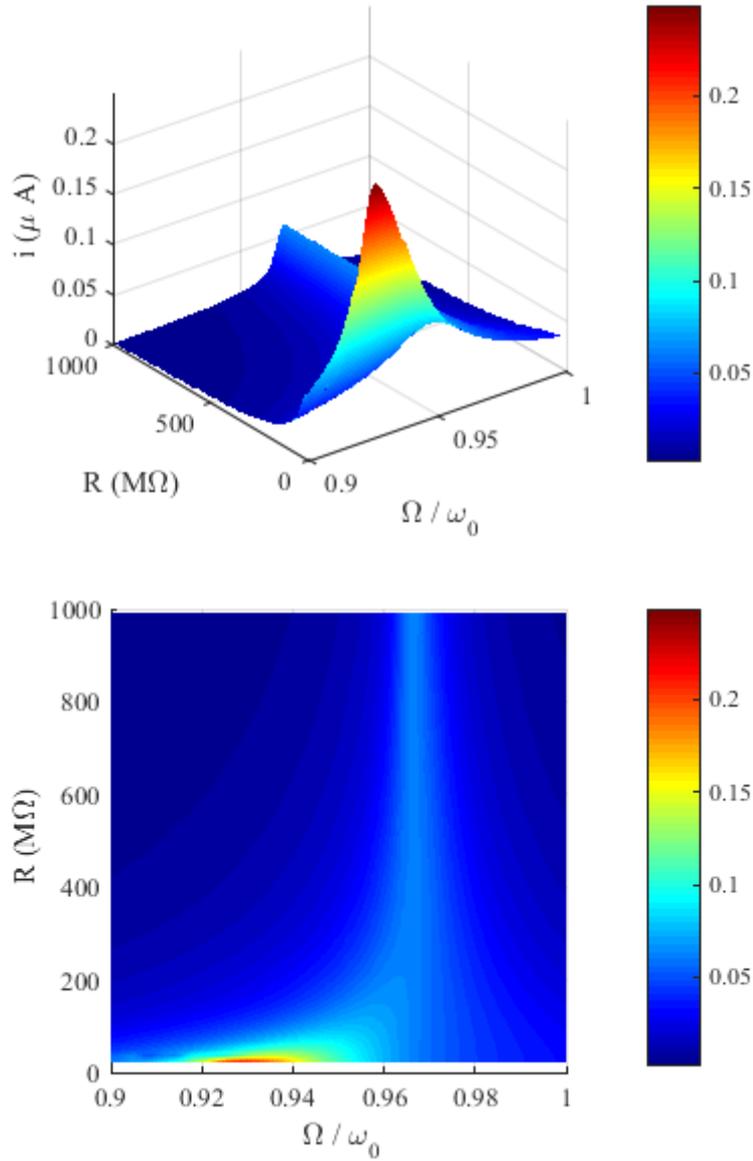

**Figure 11.** Maximum electric current as a function of excitation frequency and resistance for $V_S = 180$ V and $W_b = 0.05$

The effect of surface voltage of the electret layer on frequency response of the micro beam and mean output power is depicted in Fig. 12 and Fig. 13, respectively. It is concluded that as the surface voltage of the electret layer decreases, the system gets stiffer and therefore the resonance frequency of the microbeam increases. By decreasing the surface voltage, the resonance amplitude of the microbeam increases and this shortens the resistance domain with which the device can work without pull-in. The minimum and maximum resistance, for which the pull-in instability does not take place, is presented in Fig. 14 as a function of electret surface voltage.

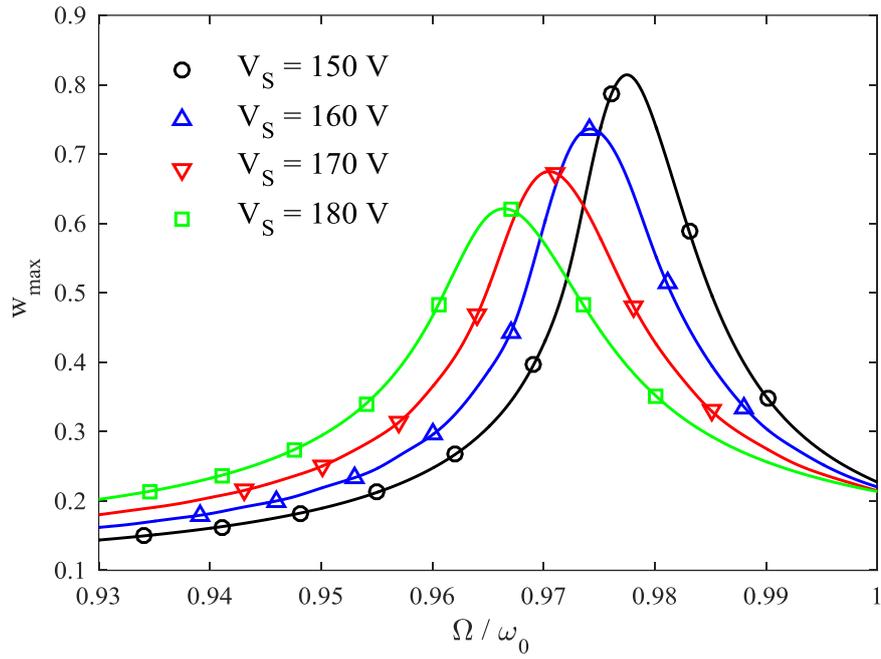

**Figure 12.** The effect of electret surface voltage on the frequency response curve of the microbeam for $R = 500$ M$\Omega$ and $W_b = 0.05$

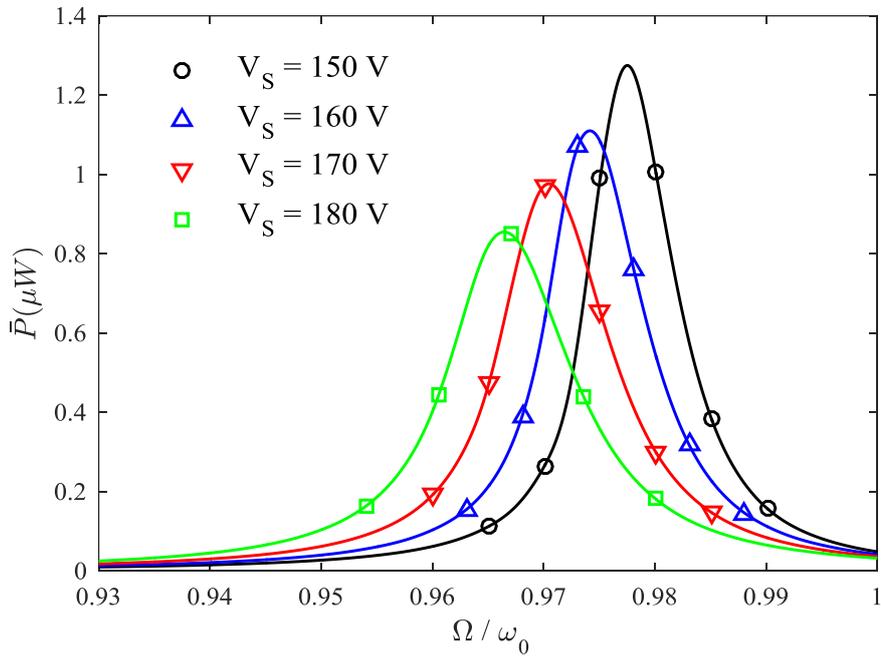

**Figure 13.** The effect of electret surface voltage on the mean output power of the system for $R = 500$ M$\Omega$ and $W_b = 0.05$

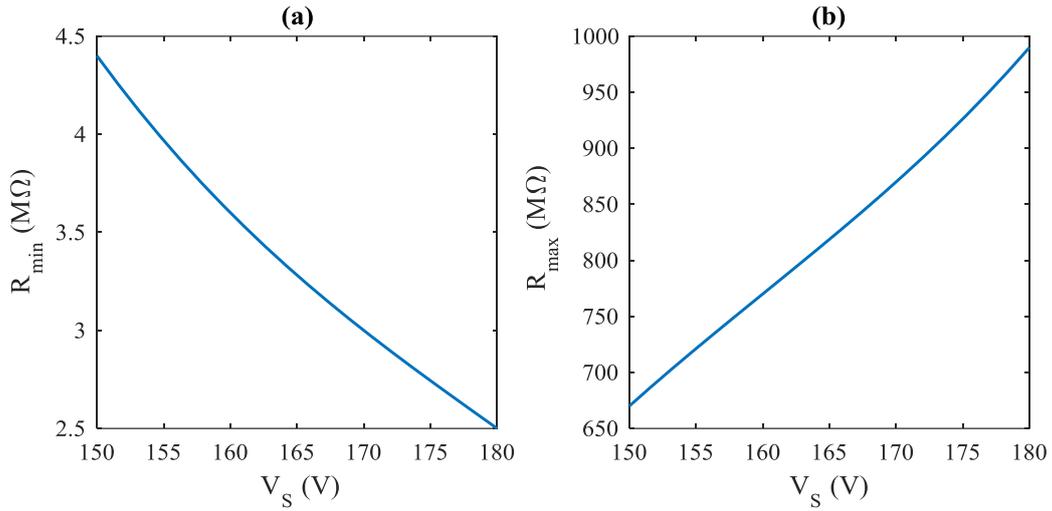

**Figure 14.** (a) Minimum and (b) maximum resistances to prevent pull-in instability as a function of surface voltage of the electret layer for $W_b = 0.05$

Fig. 15 and Fig. 16 illustrates the effect of excitation amplitude on the device performance. By increasing the excitation amplitude, the deflection and hence the output power of the system increases, however, increasing the excitation amplitude reduces the working resistance domain of the energy harvester.

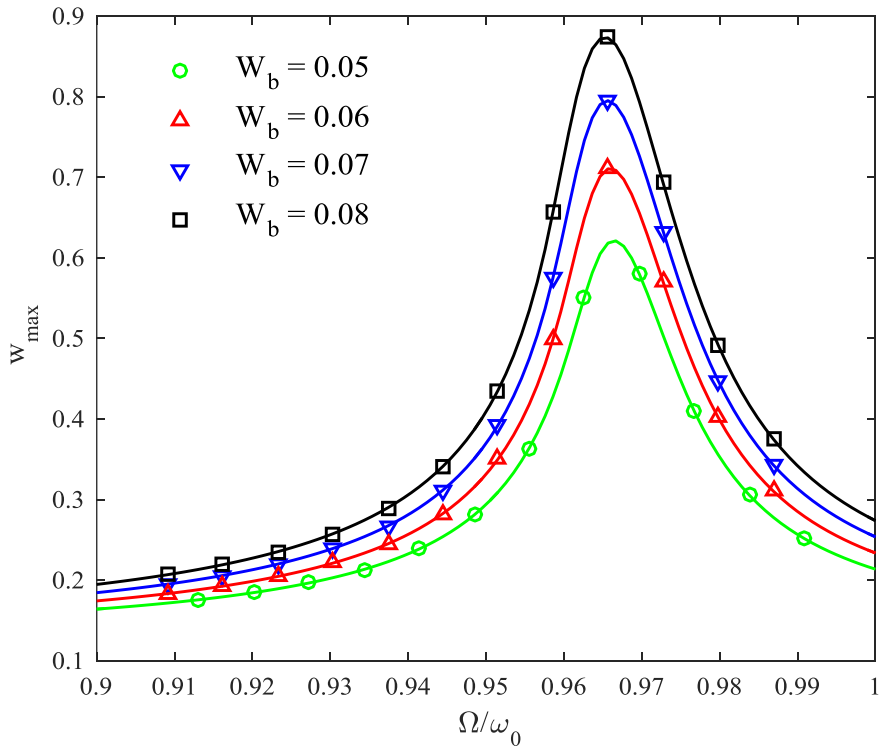

**Figure 15.** The effect base excitation amplitude on the frequency response curve of the microbeam for $R = 5 \times 10^8$ Ω and $V_S = 180$ V

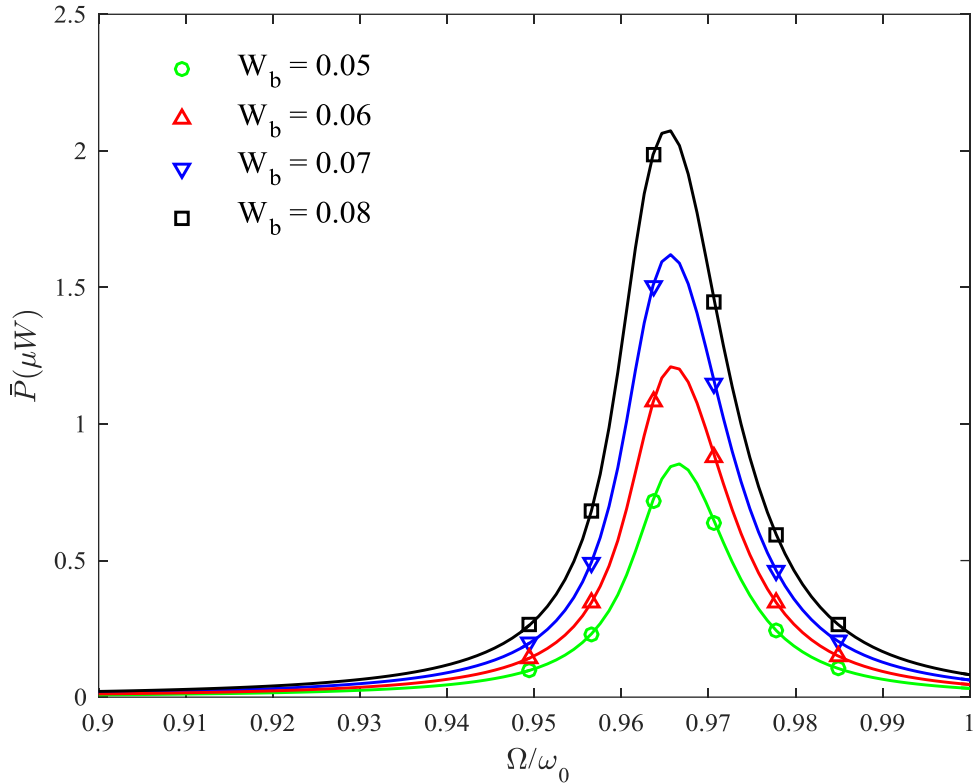

**Figure 16.** The effect base excitation amplitude on the mean output power of the system for $R = 5 \times 10^8\ \Omega$ and $V_S = 180$ V

## 5. CONCLUSION

A novel capacitive energy harvester is developed in this paper. The system is made up of a micro cantilever and a fixed substrate that form a variable capacitor together. An electret layer is attached to the substrate which provides the required bias voltage of the device. The variable capacitor is in series with a resistance and the whole system mounted on a package that the ambient vibration is applied to it as a harmonic excitation. Unlike the previous studies, the resonant system is modeled continuously based on linear Euler-Bernoulli beam theory. The governing mechanical equation of motion is derived using Hamilton's principle. The accuracy of the electro-mechanical model is verified by performing an energy conservation study in free oscillations of the system; based on the results in this case, the resistance has a damping effect the on the free vibration response of the microbeam and hence, the equivalent damping coefficient of the corresponding resistances which resembles the output power, are determined. The frequency response curve and the mean output power of the device are obtained in the presence of ambient vibrations and the effect of various parameters including resistance, electret surface voltage and the frequency and amplitude of base excitation, is investigated on the energy harvester's performance. Results show that the resistance has a significant effect on the response of the microbeam, so in low resistances, the microbeam has a softening behavior, however, by increasing the resistance the microbeam gets stiffer and represents a linear behavior. In a specific range of resistance and excitation frequency, a theoretical mean output power in the order of 1 microwatt is obtained.